\documentclass[twocolumn,showpacs,preprintnumbers,amsmath,amssymb]{revtex4}

\usepackage{graphicx}	
\usepackage{dcolumn}	
\usepackage{amssymb}

\begin{document}

\title{Magnetic Field Dependence of Spin Glass Free Energy Barriers}

\author{Samaresh~Guchhait$^1$}

\author{Raymond~L.~Orbach$^2$}

\affiliation{$^{1}$Microelectronics Research Center, The University of Texas at Austin, Austin, Texas 78758}

\affiliation{$^{2}$Texas Materials Institute,~The University of Texas at Austin,~Austin, Texas 78712}

\date{\today}   

\begin{abstract}
\noindent 
We measure the field dependence of spin glass free energy barriers in a thin amorphous Ge:Mn film through the time dependence of the magnetization.  After the correlation length $\xi(t, T)$ has reached the film thickness $\mathcal {L}=155$~\AA~so that the dynamics are activated, we change   the initial magnetic field by $\delta H$.  In agreement with the scaling behavior exhibited in a companion Letter [Janus collaboration: M. Baity-Jesi {\it et al.}, Phys. Rev. Lett. {\bf 118}, 157202 (2017)], we find the activation energy is increased when $\delta H < 0$.  The change is proportional to $(\delta H)^2$ with the addition of a small $(\delta H)^4$ term.  The magnitude of the change of the spin glass free energy barriers is in near quantitative agreement with the prediction of a barrier model. 
\end{abstract}



\maketitle



{\it Introduction.\hspace{5mm}} 
The effect of a magnetic field on spin glass dynamics has been a source of controversy for almost twenty five years.  Mean field solutions lead to a phase transition in the presence of a magnetic field, the de Almeida-Thouless transition [1], while the droplet model [2-6] predicts the vanishing of the spin glass state no matter how small the magnetic field.  Though the two are contradictory, they are difficult to distinguish experimentally [7-10].  For example, both predict a length scale, $\mathcal {L}$, dependent maximum barrier height, but with differing dependence upon $\mathcal {L}$ (see below).  In addition, both predict a decrease in effective waiting times [11,12] proportional to the square of the magnetic field strength (Refs. [8] and [13], respectively).  The present Letter probes the nature of these dynamics in the presence of a magnetic field.

The study of spin glass dynamics, especially in reduced dimensions, provides a window into the slow response of disordered and glassy systems [14].  Further, the approach to critical behavior has analogies with structural glasses [15].  Characteristic times for spin glass response can vary from laboratory time scales to impossibly long times as a consequence of highly degenerate states well separated in phase space [16].  

This Letter reports measurements of the effect of magnetic field changes on the free energy barriers in thin film spin glasses, where the correlation length, $\xi(t, T)$ at time $t$ and temperature $T$ is confined by $\xi(t, T)\leq \mathcal {L}$, the film thickness.  
The number of participating spins $N$ is order of $\sim\!({\mathcal {L}}/{a_0})^3$, where $a_0$ is the average distance between spins.     
Our results demonstrate experimentally that spin glass free energy barriers are affected as the square of magnetic field changes, consistent with [8,13], plus a small fourth order term. 
An   accompanying Letter, Baity-Jesi {\it et al.}~[17], using numerical   simulations on {\sl Janus II} arrives at equivalent conclusions.
 The magnitude of the effect is consistent with a {\it barrier model} estimate based on the observed magnetization.  A {\it trap model} [18,19] would predict a change in barrier heights {\it linear} in the change of magnetic field.  However, at the fields used in our experiments, it is found to be too small by around two orders of magnitude from that which we observe.

Spin glass dynamics at the mesoscale (length scales $1\leq {\mathcal {L}}\leq 100$ nm) [20] are achievable in thin film multilayers [21-24], and have been reported for a Ge:Mn thin film~[25]. The beauty of thin film spin glasses with mesoscopic thickness $\mathcal {L}$ lies with the growth on laboratory time scales of the spin glass correlation length $\xi(t, T)$ from $\xi(t, T)\leq {\mathcal {L}}$ to $\xi(t_{\text {co}}, T)={\mathcal {L}}$ for $t < t_{\text {co}}$ to $t = t_{\text {co}}$,  defining the crossover time $t_{\text {co}}$.  After $\xi(t, T)$ reaches $\mathcal {L}$, there is no further growth of $\xi(t, T)$ at fixed temperature.

The growth of $\xi(t, T)$ from nucleation is different between the two models introduced in the first paragraph.  The droplet model~[2-6] assumes activated growth and finds,
\begin{equation}
\xi(t, T)=\alpha\,a_0\left[\left({\frac {T}{T_g}}\right)\ln\left({\frac {t}{\tau_0}}\right)\right]^{1/\psi},
\end{equation}
where $\alpha$ is a normalization constant of order unity, $\tau_0$ is an exchange time of the order of $\hbar/(k_BT_g)$, and $\psi$ is a critical exponent.  Experiments [26-28] and simulations [29,30] find values of $\psi$ between 0.65 and 1.1, with most values close to unity.  The spin glass dynamics are activated when $\xi(t, T)=\mathcal{L}$ with the largest activation energy,
\begin{equation}
\frac{\Delta_{\text {max}}(\mathcal {L})}{k_BT_g}=\left({\frac {\mathcal {L}}{\alpha a_0}}\right)^{\psi}.
\end{equation}

The model based on the mean field solution [31-33] uses a power law growth for $\xi(t, T)$, 
\begin{equation}
\xi(t, T)=c_1\,a_0\left({\frac {t}{\tau_0}}\right)^{c_2(T/T_g)},
\end{equation}
where $c_1$ and $c_2$ are constants determined from experiment.  The dynamics are also activated when $\xi(t, T)=\mathcal{L}$ with the largest  activation energy,
\begin{equation}
\frac {\Delta_{\text {max}}(\mathcal{L})}{k_BT_g}={\frac {1}{c_2}}\left[\ln\left({\frac {\mathcal{L}}{a_0}}\right)-\ln\,c_1\right].
\end{equation}
Thus, {\it both} models predict activated dynamics when $\xi(t, T)=\mathcal{L}$, but with differing dependence upon the length scale $\mathcal{L}$.  This Letter will not attempt a choice between models.  It suffices to say that what will be reported is the magnetic field dependence of the largest free energy barrier, the  activation energy $\Delta_{\text {max}}(\mathcal {L})$.

The actual measurements of spin glass dynamics require the presence of a magnetic field $H$.  If the spin glass is rapidly cooled from above the spin glass transition temperature $T_g$ to a quench temperature $T_q$ below $T_g$ in zero magnetic field, dynamics are generated upon the application of a magnetic field through measurement of the zero-field-cooled (ZFC) magnetization  $M_{\text {ZFC}}(t, H)$.  If the spin glass is cooled in a magnetic field from above $T_g$ to $T_q$, and the magnetic field is reduced to zero, the measured magnetization is the thermoremanent magnetization (TRM) and termed $M_{\text {TRM}}(t, H)$.  We are omitting discussion of the {\it waiting time} effect [11,12] because it is irrelevant as long as $t\geq t_{\text {co}}$.  We exhibit experimental results below of magnetic field changes upon $\Delta_{\text {max}}(\mathcal{L})$ for a thin film Ge:Mn (155~\AA) spin glass.  After that we analyze the experimental results in terms of the power law dependence of $\delta \Delta_{\text {max}}$ on the change of magnetic field $\delta H$, and then follow by a summary.\\

{\it Experimental results.\hspace{5mm}} 
The experiments were performed on thin amorphous Ge:Mn (11~at.\% Mn) films of thickness 155~\AA~with a glass temperature $T_g \approx 24$~K~[34].     Previous experiments have shown this insulating system to exhibit spin glass properties [34,35], not unlike Eu$_x$Sr$_{1-x}$S~[36], an  insulating canonical spin glass system.  Further, the behavior of the field-cooled (FC) magnetization is very similar to that found for insulating Eu$_x$Sr$_{1-x}$S~[36] and the thinnest Cu:Mn films by G. G. Kenning {\it et al.}~[21].   All the dynamical measurements~[25] on these films  are consistent with the  usual spin glass systems, establishing confidence in the generality of effects seen in Ge:Mn films.

\begin{figure}[t]
\vspace{-0mm}
\includegraphics[width=8.6cm]{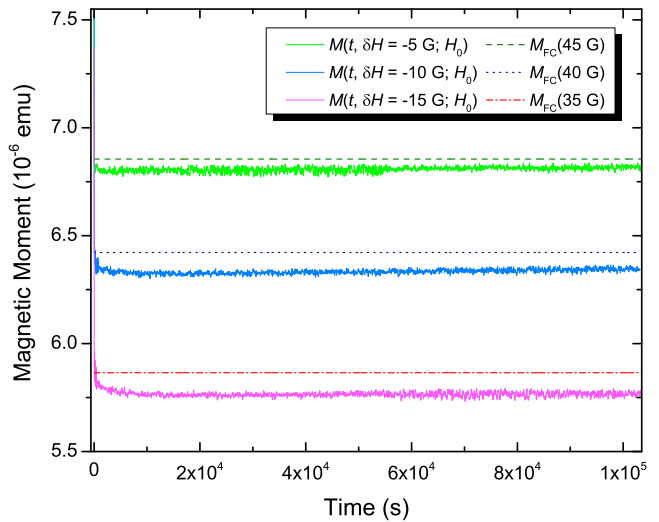}
\vspace{-6mm}
\caption{Magnetization measurements, $M(t,\delta H; H_0)$, after zero-field cooling at 21.5 K, and reducing the original magnetic field $H_0=50$ G after twenty hours by representative values $\delta H$ =  -5, -10, and -15 G, as a funtion of time.  Also plotted is the field-cooled magnetization, $M_{\text {FC}}(H_0+\delta H)$, to which the measured moment is approaching.}
\label{fig:fig1}
\end{figure}

This amorphous Ge:Mn  thin film sample was quenched to a temperature $T_q = 21.5$~K in zero magnetic field.  The quench temperature $T_q$ was chosen so that $\xi(t, T)$ could grow to the thickness of the sample, $\mathcal {L}$, on a reasonable laboratory time scale.  Previous measurements [25] found this crossover time to be about $t_{\text {co}}\approx 6.8\times 10^4$~sec, or about 19 hours. In our experiments, after the temperature is stabilized at $T_q$, a magnetic field $H_0=50$~G is applied, and the system allowed to age for 20 hours.  This ensures that the correlation length has reached the sample thickness.

During this aging period, the zero-field cooled magnetization, $M_{\text {ZFC}}(t)$, increases, but the increase is sufficiently slow that it remains well below the field-cooled  value, $M_{\text {FC}}$, on this time scale.  The slope of the irreversible component of the magnetization, $M_{\text {FC}}-M_{\text {ZFC}}(t\geq t_{\text {co}})$, yields $\Delta_{\text {max}}=37.5~k_BT_g$, as before~[25].

The experiment was repeated exactly as above, but the applied magnetic field ($H_0$ = 50~G) was reduced by $\delta H$ {\it after} 20 hours.  The subsequent measured magnetization $M(t,\delta H; H_0)$ is exhibited in Fig.~1 as a function of time for the representative values $\delta H$ = -5, -10, and -15~G.  
In all, experiments using this protocol were performed with $\delta H$ = -5, -10, -12.5, -15, and -17.5~G.  Activated behavior is seen in all, with $\Delta_{\text {max}}(\mathcal{L})$ increasing as the magnetic field is reduced.  The irreversible part of the magnetization, $M_{\text {FC}}(H_0+\delta H) - M(t, \delta H; H_0)$ is plotted in semi-log scale in Fig.~2 for the representative values $\delta H$ = -5, -10, and -15~G.

\begin{figure}[b]
\vspace{-0mm}
\includegraphics[width=8.6cm]{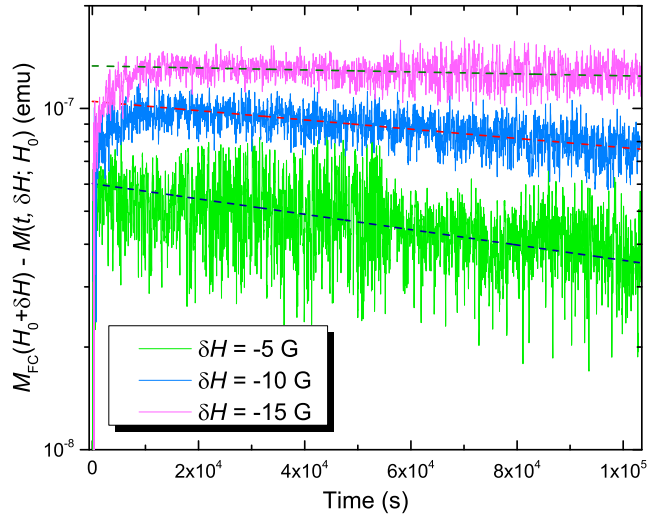}
\vspace{-6mm}
\caption{The logarithm of the irreversible part of magnetization, $\log_{10}[M_{\text {FC}}(H_0+\delta H)-M(t,\delta H; H_0)$], for the representative values $\delta H$ = -5, -10, and -15 G, plotted against the time.  The dashed straight lines, displaying activated behavior, give rise to the values of $\Delta_{\text {max}}(H_0+\delta H)$ displayed in Fig.~3.}
\label{fig:fig2}
\end{figure}

Figs.~1 and 2 display a curious behavior at short times after the magnetic field change.  From Fig.~1, $M(t, \delta H; H_0)$ initially {\it decreases} with time until about $\sim$20,000 sec, when it turns around and begins to increase with time, as expected for $M_{\text {ZFC}}(t)$.  This is mirrored in Fig.~2 with an initial rise in $M_{\text {FC}}(H_0+\delta H)-M(t, \delta H; H_0)$ out to $\sim$20,000 sec, after which activated decay is seen.  The system appears to be increasing its irreversible magnetization initially after the magnetic field has changed.   
The {\it curing time} for a return to activated decay for the irreversible behavior is roughly the same as that found for temperature chaos [25].  Fig.~2 suggests that this curing time is independent of the magnitude of the reduction in magnetic field.

For times greater than  $\sim\!$ 20,000~sec, the decay of $M_{\text {FC}}(H_0+\delta H)-M(t, \delta H; H_0)$ is activated (Fig.~2).  The slopes of the decay curves for each value of $\delta H$ generate values for $\Delta_{\text {max}}(H_0+\delta H)$~[25]. 
Fig.~3 plots $\delta\Delta_{\text {max}}(\delta H) = \Delta_{\text {max}}(H_0+\delta H) - \Delta_{\text {max}}(H_0)$  against $\delta H$.  
Two dependences, $\delta\Delta_{\text {max}}(\delta H) \propto \delta H$ and $\delta\Delta_{\text {max}}(\delta H)\propto (\delta H)^2$, are plotted along with the data.  It is clear from Fig.~3 that $\delta \Delta_{\text {max}}$ varies more rapidly than linear in $\delta H$.  
On the assumption of higher-order non-linearity in $\delta\Delta_{\text {max}}(\delta H)$, we have added a small   $(\delta H)^4$ term to the quadratic fit~[17].  The points can also be fit with an analytic form $a_1(\delta H)^2/\sqrt {1+{a_2}(\delta H)^2}$~[37]. The values of the parameters used for these fitting procedures are given in Table~I.

The $\chi^2$ {\sl goodness-of-fit  probability}, $Q$~[38], is also listed in Table~I for each of the four proposed dependences.  Not surprisingly, as seen from Fig.~3, it is very small for $\delta \Delta_{\text {max}}(\delta H)\propto \delta H$.  Likewise, it is finite but small for $\delta\Delta_{\text {max}}(\delta H)\propto (\delta H)^2$.  The inclusion of the fourth order term is convincing, with $Q \approx 0.88$, though the additional fitting parameter of course reduces the degrees of freedom by one.  The analytic fit to the dependence $a_1(\delta H)^2/\sqrt {1+{a_2}(\delta H)^2}$ has a slightly smaller probability, $Q \approx 0.77$, but is an arbitrary form.  The sum and substance of Fig.~3 and Table~I is simply that the data do not fit a linear dependence of  $\delta\Delta_{\text {max}}(\delta H)$ on $\delta H$, but rather  a quadratic relationship with a small fourth order term.

Previous experiments~[8]  have used the magnetic field variation of the effective waiting time to determine the magnetic field dependence of $\Delta_{\text {max}}$.  They posit a reduction in $\Delta_{\text {max}}$ through the magnitude of the Zeeman energy $E_{\text {Z}}(H)$ from an {\sl effective waiting time} $t_w^\text{eff}$ through [their Eq.~(4)],
\begin{equation}
\Delta_{\text{max}}(t_w, T)-E_{\text {Z}}(H)=k_BT\left(\ln\,t_w^{\text {eff}} - \ln\,\tau_0\right).
\end{equation}
\noindent
Their experiments were conducted on bulk samples with a concomitant distribution of length scales $\mathcal{L}$, and therefore exhibit an average over a distribution of $\Delta_{\text {max}}(\mathcal{L})$~[24].  Further, the correlation length continues to increase in time in their experiments, requiring the use of an effective waiting time to extract the magnetic field dependence of the barrier heights.   Fig.~2 in Ref.~[8] exhibits a quadratic dependence of $\log t_w^{\text{eff}}$ vs $H^2$ for small $H$, and hence a quadratic reduction in barrier heights $\Delta$ with magnetic field.  
This is equivalent to our extraction of $\Delta_{\text {max}}$ with $(\delta H)^2$. 
However, our experiments are conducted on a mesoscopic thin film with a single thickness $\mathcal {L}$.  As a consequence, the correlation length growth terminates at the length scale $\mathcal {L}$. There is no averaging in our experiments: $\Delta_{\text {max}}(\mathcal {L})$ is set by the thin film thickness, enabling a very accurate determination of the values for $\delta \Delta_{\text {max}}$ as a function of magnetic field change. These differences distinguish the present set of experiments from those in Ref.~[8].\\

\begin{figure}
\vspace{-0mm}
\includegraphics[width=8.6cm]{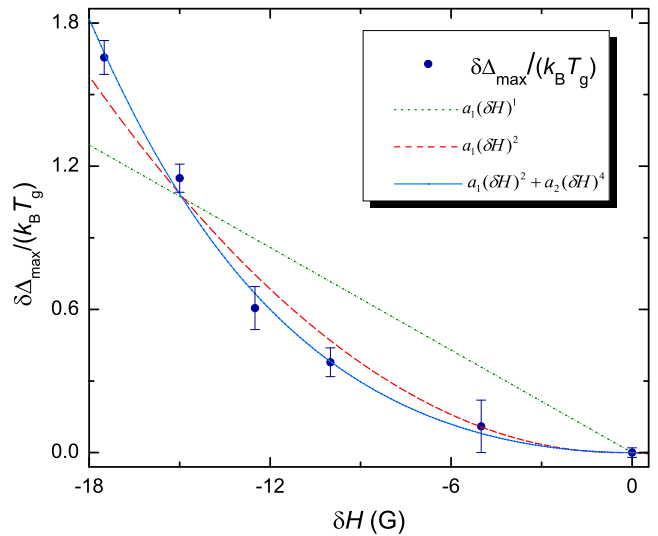}
\vspace{-6mm}
\caption{Plot of the measured increases of the maximum barrier height, $\Delta_{\text {max}}(H_0+\delta H)$ as a function of the reductions in magnetic field $\delta H$.  Shown on the figure are curves for the fit to a linear variation in $\delta H$ (dotted line), a quadratic variation in $\delta H$ (dashed line),  and with the addition of a small fourth order term to the quadratic variation (solid line).  The numerical fitting values are given in Table~I.}
\label{fig:fig3}
\end{figure}

\begin{table*}
    \begin{tabular}{| l | l | l | l | l | l |}
    \hline
    Fit & $a_1$ & $a_2$ & RMS Error & R-squared &$\chi^2$ goodness-of-fit, $Q$ \\ \hline
    \hline
    $a_1 (\delta H)^1$ &  $-7.16\times10^{-2}$~G$^{-1}$ & &0.291 & 0.793  & $2 \times10^{-16}$ \\ \hline
    $a_1(\delta H)^2$ &  $5.00\times10^{-3}$~G$^{-2}$ & & 0.111 & 0.970  & 0.0663 \\ \hline
    $a_1(\delta H)^2+a_2(\delta H)^4$ &  $3.01\times10^{-3}$~G$^{-2}$ & $8.03\times10^{-6}$~G$^{-4}$ & 0.048  & 0.995  & 0.8787 \\ \hline
    $\frac{a_1(\delta H)^2}{\sqrt{1+a_2(\delta H)^2}} $&  $3.67\times10^{-3}$~G$^{-2}$ & $-1.80\times10^{-3}$~G$^{-2}$ & 0.056  & 0.994   & 0.7686 \\
    \hline
    \end{tabular}
    \caption{Parameters used for the best fits to the data exhibited in Fig.~3.  There are two higher-order non-linear forms, indistinguishable from one another on Fig.~3.  One is a simple sum of quadratic and quartic terms; the other is suggested by the analysis in the accompanying Letter~[17]. The  uncertainties of $\Delta_\text{max}$ in Fig.~3 are used as estimates of the standard deviations for each of the measurements to calculate $Q$~[38].}
\end{table*}


{\it Analysis of experimental results.\hspace{5mm}} 
Because the lower critical dimension for spin glasses, $d_\ell \approx 2.5$~[39-41], a spin glass at dimension $d = 3$ will exhibit a finite glass transition temperature $T_g$, while a spin glass at dimension $d = 2$ will have $T_g = 0$.  As a consequence, as outlined in Ref.~[24], the spin glass correlation length will be anisotropic at $t>t_{\text {co}}$.  The component perpendicular to the film layer will saturate at $\xi_\perp=\mathcal {L}$, while the parallel component, $\xi_\parallel$, experiences $d=2$ critical fluctuations.  A scaling form [37], built on the assumption of multiplicative growth consistent with Ref.~[24], suggests that $\xi_{\parallel}(T)$ saturates at 
\begin{equation}
\xi_{\parallel}(T) =k(T)\,\mathcal {L}=b\,(T_g/T)^{\nu_{\text{2d}}}\mathcal {L}.  
\end{equation}
Here, $b$ is a constant of order unity, and $\nu_{\text{2d}}$ is the usual $d=2$ critical exponent, $\nu_{\text{2d}}\approx 3.53(7)$~[42].  The correlated spins thus have a {\sl pancake-like} structure, with the parallel dimension larger than the perpendicular dimension, the former increasing with decreasing temperature.  
The volume of the correlated spins is that of the {\it pancake} described above, encapsulated in the perpendicular direction by $\mathcal {L}$ and in the parallel direction by the area, $\pi [\xi_{\parallel}(T_q)]^2=\pi {b^2}{\mathcal {L}^2}(T_g/T_q)^{2\nu_{\text{2d}}}$. 

It is interesting to investigate the magnitude of the two predictions of the variation of $\Delta_{\text {max}}$ with magnetic field change. 
The {\it trap model} [18,19] ``associates a typical Zeeman energy $E(\Delta H)$ to the field variation $\Delta H$; ...." for the reduction of the effective trap depth.  They take,
\begin{equation}
\delta E_{\text {Z}}(\delta H) \equiv \delta E_{\text {Z}}(N, \delta H)=m\,\mu_B\,{\sqrt {N}}\,\delta H.
\end{equation}
The magnetic moment $M=m\mu_B$ refers to ``single spins, but also renormalized groups of spins."  Because of the ``random nature of the interactions and the frustration they cause, the net uncompensated moment for a group of $N~spins$ is of the order of $\sqrt {N}$, ....".  Thus, the trap model predicts a linear relationship between the change in the Zeeman energy and the change in magnetic field.

The number of spins, $N$, contained within the correlated volume in the Ge:Mn 11 at.\% 155~\AA~film $(a_0 $ = 5.3~\AA) is approximately 170,840, so that $\sqrt {N}\approx 413$.  Taking $m\sim 1$, Eq.~(7) reduces to  $\delta E_Z(\delta H)\sim0.3$~K for $\delta H= -10$~G.   From Fig.~3, $\delta E_Z(\delta H=-10~{\text {G}})\sim 0.4~T_g\sim 10$~K, so that Eq.~(7) is too small by around two orders of magnitude.  However, our experiments were performed in significant magnetic fields.  The fluctuation result of Eq.~(7) would surely be relevant in experiments carried out in zero or very small magnetic fields.

Use of Eq.~(5)  reduces the values of the free energy barrier $\Delta$ uniformly by the change in the Zeeman energy, $E_{\text {Z}}(H)$.  The difference for mesoscopic systems is that the reduction is {\it independent} of time for $t\geq t_{\text {co}}$.  That is, in an obvious notation, $\Delta_{\text {max}}(\mathcal{L},H_0+\delta H)$ is given by:
\begin{equation}
\Delta_{\text {max}}(\mathcal{L},H_0+\delta H)=\Delta_{\text {max}}(\mathcal{L},H_0)-\delta E_{\text {Z}}(\,\delta H)
\end{equation}
and we measure $\Delta_{\text {max}}({\mathcal {L}}, H_0+\delta H)$ directly as described above.  
The  {\it barrier model}~[8] sets
\begin{equation}
E_{\text {Z}}(H)=N\chi_{_{\text{FC}}} H^2,
\end{equation}
where $N$ ``... defines a volume over which the spins are effectively locked together for barrier hopping, the radius of which we define as the spin glass correlation length $\xi(t, T)$" [8].  
In our case, we take the volume to be $\pi [\xi_{\parallel}(T_q)]^2\,\xi_{\perp}=\pi {b^2}{\mathcal {L}^3}(T_g/T_q)^{2\nu_{\text{2d}}}$ as before.  
The field cooled  magnetic susceptibility per spin in [8] was taken as $\chi_{_{\text {FC}}}$ because they were measuring the time dependence of the thermoremanent magnetization, $M_{\text {TRM}}(t)$.  In our case it is the zero-field-cooled magnetization, $M_{\text {ZFC}}(t)$.  Given how close $M_{\text {ZFC}}(t)$ is to $M_{\text {FC}}$ in our experiments, the difference is negligible.

From Fig.~1, the field-cooled magnetic moment $M_{\text {FC}} \approx 6.4\times 10^{-6}$~emu for $H=40$~G.  Using Eq.~(9),  we get [24],
\begin{eqnarray}
\delta E_Z &=& 2N\chi_{_{\text {FC}}}H\,\delta H = 2N\left(\frac {M_{\text {FC}}}{N_{\text {t}}}\right)\,\delta H  \nonumber \\
&\approx&  2M_{\text{FC}}\left[\pi b^2 \mathcal{L}^3 \left({\frac {T_g}{T_q}}\right)^{2\nu_{\text{2d}}}\!\!
{\Big/}V_{\text {s}}\right]\,\delta H,
\end{eqnarray}
where $N_{\text {t}}$ is the total number of spins in the sample and $V_{\text {s}}=2.06\times 10^{-6}$~cm$^3$, the total volume of the sample, using the known thickness and estimated area. Using    $\mathcal {L}=15.5$~nm, $b$ equal to unity, $T_g=24$~K, $T_q=21.5$~K, $\nu_{\text{2d}}=3.53$,  we find $\delta E_{\text {Z}}\sim 11.4$~K for $\delta H=-10$~G, very close to the measured value for $\delta E_{\text {Z}}(\delta H=-10$~G) $\sim 10$~K in Fig.~3.\\

{\it Summary.\hspace{5mm}} 
The reduction in the free energy barrier height responsible for spin glass dynamics at the mesoscale is measured as a function of magnetic field change in a 155 \AA~ Ge:Mn 11 at.\% thin film.  It is found that the magnitude of the reduction varies as the square of the change in magnetic field, $(\delta H)^2$ with a small $(\delta H)^4$ term.  This result is consistent with the scaling laws of a companion Letter [17].  Quantitative estimates of two prevalent models are also presented.  The magnitude of the prediction of a trap model appears to be too small to fit the data by nearly two orders of magnitude for the magnetic fields used in these experiments.  A barrier model predicated on a change of the maximum barrier height $\Delta_{\text {max}}(\mathcal{L}, H)$ with magnetic field agrees with the measurements, both in terms of the relationship to the change in magnetic field, and nearly quantitatively with its magnitude. \\

{\it Acknowledgments.\hspace{5mm}} 
We thank the Janus collaboration for sharing with us their results prior to publication.  In addition, as part of that collaboration, the authors express deep gratitude to Dr. Victor Martin-Mayor for his assistance in the analysis presented in this paper.  This work was supported by the U.S. Department of Energy, Basic Energy Sciences, Award No. DE-SC0013599.\\
\\
\centerline {\bf References}
\centerline{}
1.  J. R. L. de Almeida and D. J. Thouless, Stability of the Sherrington-Kirkpatrick Solution of a Spin Glass Model, J. Phys. A: Math. Gen. {\bf 11}, 129 (1978).\\
2.  D. S. Fisher and D. A. Huse, Ordered Phase of Short-Range Ising Spin-Glass, Phys. Rev. Lett. {\bf 56}, 1601 (1986).\\
3.  D. A. Huse and D. S. Fisher, Pure States in Spin Glasses, J. Phys. A: Math. Gen. {\bf 20}, L997 (1987).\\
4.  D. S. Fisher and D. A. Huse, Static and Dynamic Behavior of Spin-Glass Films, Phys. Rev. B {\bf 36}, 8937 (1987).\\
5.  D. S. Fisher and D. A. Huse, Nonequilibrium Dynamics of Spin Glasses, Phys. Rev. B {\bf 38}, 373 (1988).\\
6.  D. S. Fisher and D. A. Huse, Equilibrium Behavior of the Spin-Glass Ordered Phase, Phys. Rev. B {\bf 38}, 386 (1988).\\
7.  F. Lefloch, J. Hammann, M. Ocio and E. Vincent, Can Aging Phenomena Discriminate between the Droplet Model and a Hierarchical Description in Spin Glasses?, Europhys. Lett. {\bf 18}, 647 (1992); K. Jonason, E. Vincent, J. Hammann, J. P. Bouchaud and P. Nordblad, Memory and Chaos Effects in Spin Glasses, Phys. Rev. Lett. {\bf 81}, 3243 (1998).\\
8.  Y. G. Joh, R. Orbach, G. G. Wood, J. Hammann and E. Vincent, Extraction of the Spin Glass Correlation Length, Phys. Rev. Lett. {\bf 82}, 438 (1999).\\
9.  K. Jonason, P. Nordblad, E. Vincent, J. Hammann and J.-P. Bouchaud, Memory Interference Effects in Spin Glasses, Eur. Phys. J. B {\bf 13}, 99 (2000).\\
10.  J.-P. Bouchaud, V. Dupuis, J. Hammann and E. Vincent, Separation of Time and Length Scales in Spin-Glasses: Temperature as a Microscope, Phys. Rev. B {\bf 65}, 024439 (2001).\\
11.  L. Lundgren, P. Svedlindh, P. Nordblad, and O. Beckman, Dynamics of the Relaxation-Time Spectrum in a CuMn Spin-Glass, Phys. Rev. Lett. {\bf 51}, 911 (1983); L. Lundgren, P. Svedlindh,  and O. Beckman, Anomalous Time-Dependence of the Susceptibility in a Cu(Mn) Spin-Glass, J. Magn. Magn. Mater. {\bf 31-34}, 1349 (1983).\\
12.  R. V. Chamberlin, Time Decay of the Thermoremanent Magnetization in Spin-Glasses as a Function of the Time Spent in the Field-Cooled State, Phys. Rev. B {\bf 30}, 5393 (1984); M. Ocio, M. Alba and J. Hammann, Time Scaling of the Aging Process in Spin-Glasses - a Study in CsNiFeF$_6$, J. Phys. Lett. (Paris), {\bf 46}, L1101 (1985).\\
13.  H. Takayama and K. Hukushima, Field-Shift Aging Protocol on 3D Ising Spin-Glass Model: Dynamical Crossover Between the Spin-Glass and Paramagnetic States, J. Phys. Soc. Japan {\bf 73}, 2077 (2004).\\
14.  L. Sandlund, P. Granberg, L. Lundgren, P. Nordblad, P. Svedlindh, J. A. Cowen and G. G. Kenning, Dynamics of Cu-Mn Spin-Glass Films, Phys. Rev. B {\bf 40}, 869 (1989).\\
15.  S. Albert, Th. Bauer, M. Michl, G. Biuroli, J.-P. Bouchaud, A. Loidl, P. Lunkenheimer, R. Tourbot, C. Wiertel-Gasquet and F. Ladieu, Fifth-Order Susceptibility Unveils Growth of Thermodynamic Amorphous Order in Glass Formers, Science {\bf 352}, 1308 (2016).\\
16.  G. Parisi, Toward a Mean Field-Theory for Spin-Glasses, Phys. Lett. {\bf 73A}, 203 (1979); Infinite Number of Order Parameters for Spin-Glasses, Phys. Rev. Lett. {\bf 43}, 1754 (1979); A Sequence of Approximated Solutions to the S-K Model for Spin-Glasses, J. Phys. A {\bf 13}, L115 (1980); M. M\'ezard, G. Parisi, N. Sourlas, G. Toulouse, and M. A. Virasoro, Replica Symmetry-Breaking and the Nature of the Spin-Glass Phase, J. Phys. (Paris) {\bf 45}, 843 (1984); M. M\'ezard and M. A. Virasoro, The Microstructure of Ultrametricity, J. Phys. (Paris) {\bf 46}, 1293 (1985).\\
17.  M. Baity-Jesi, E. Calore, A. Cruz, L. A. Fernandez, J. M. Gil-Narvion, A. Gordillo-Guerrero, D. I\~niguez, A. Maiorano, E. Marinari, V. Martin-Mayor, J. Monforte-Garcia, A. Mu\~noz-Sudupe, D. Navarro, G. Parisi, S. Perez-Gaviro, F. Ricci-Tersenghi, J. J. Ruiz-Lorenzo, S. F. Schifano, B. Seoane, A. Tarancon, R. Tripiccione and D. Yllanes, Matching Microscopic and Macroscopic Responses in Glasses, Phys. Rev. Lett. {\bf 118}, 157202 (2017).\\ 
18.  J.-P. Bouchaud, Weak Ergodicity Breaking and Aging in Disordered Systems, J. Phys. {\bf I}, {\bf 2}, 1705 (1992).\\
19.  E. Vincent, J.-P. Bouchaud, D. S. Dean and J. Hammann, Aging in spin glasses as a random walk: Effect of a magnetic field, Phys. Rev. B {\bf 52}, 1050 (1995-II).\\
20.  {\texttt https://science.energy.gov/$\sim$/media/bes/pdf/\\reports/files/From\_Quanta\_to\_the\_Continuum\_rpt.pdf} \\
21.  G. G. Kenning, J. Bass, W. P. Pratt, Jr., D. Leslie-Pelecky, W. Leach, M. L. Wilson, R. Stubi and J. A. Cowan, Finite Size Effects in Cu-Mn Spin Glasses, Phys. Rev. B {\bf 42}, 2393 (1990).\\
22.  L. Sandlund, P. Granberg, L. Lundgren, P. Nordblad, P. Svedlindh, J. A. Cowan and G. G. Kenning, Dynamics of Cu-Mn Spin-Glass Films, Phys. Rev. B {\bf 40}, 869 (1989).\\
23.  P. Granberg, P. Nordblad, P. Svedlindh, L. Lundgren, R. Stubi, G. G. Kenning, D. L. Leslie-Pelecky, J. Bass and J. Cowan, Dimensionality Crossover in CuMn Spin-Glass Films, J. Appl. Phys. {\bf 67}, 5252 (1990).\\
24.  S. Guchhait, G. G. Kenning, R. Orbach, and G. F. Rodriguez, Spin Glass dynamics at the mesoscale, Phys. Rev. B {\bf 91}, 014434 (2015).\\
25.  S. Guchhait and R. Orbach, Direct dynamical evidence for the spin glass lower critical dimension $2<d_{\ell}<3$, Phys. Rev. Lett. {\bf 112}, 126401 (2014); S. Guchhait and R. L. Orbach, Temperature chaos in a Ge:Mn thin-film spin glass, Phys. Rev. B {\bf 92}, 214418 (2015).\\
26.  A. G. Schins, A. F. M. Arts and H. W. de Wijn, Domain Growth by Aging in Nonequilibrium Two-Dimensional Random Ising Systems, Phys. Rev. Lett. {\bf 70}, 2340 (1993).\\
27.  C. Dekker, A. F. M. Arts, H. W. de Wijn, A. J. van Duyneveldt and J. A. Mydosh, Activated Dynamics in the Two-Dimensional Ising Spin-Glass Rb$_2$Cu$_{1-x}$Co$_x$F$_4$, Phys. Rev. Lett. {\bf 61}, 1780 (1988).\\
28.  C. Dekker, A. F. M. Arts, H. W. de Wijn, A. J. van Duyneveldt and J. A. Mydosh, Activated Dynamics in a Two-Dimensional Ising Spin Glass: Rb$_2$Cu$_{1-x}$Co$_x$F$_4$, Phys. Rev. B {\bf 40}, 11 243 (1989).\\
29.  T. R. Gawron, M. Cieplak and J. R. Banavar, Scaling of Energy Barriers in Ising Spin-Glasses, J. Phys. A {\bf 24}, L127 (1991).\\
30.  H. Rieger, B. Steckemetz and M. Schreckenberg, Aging and Domain Growth in the Two-Dimensional Ising Spin Glass Model, Europhys. Lett. {\bf 27}, 485 (1994).\\
31.  E. Marinari, G. Parisi, J. Ruiz-Lorenzo, and F. Ritort, Numerical Evidence for Spontaneously Broken Replica Symmetry in 3D Spin Glasses, Phys. Rev. Lett. {\bf 76}, 843 (1996).\\
32.  J. Kisker, L. Santen, M. Schreckenberg and H. Rieger, Off-Equilibrium Dynamics in Finite-Dimensional Spin-Glass Models, Phys. Rev. B {\bf 53}, 6418 (1996).\\
33.  J.-O. Andersson and P. Sibani, Domain Growth and Thermal Relaxation in Spin Glasses, Physica (Amsterdam) {\bf 229A}, 259 (1996).\\
34.  S. Guchhait, M. Jamil, H. Ohldag, A. Mehta, E. Arenholz, G. Lian, A. LiFatou, D. A. Ferrer, J. T. Markert, L. Colombo and S. K. Banerjee, Ferromagnetism in Mn-implanted epitaxially
grown Ge on Si(100), Phys. Rev. B {\bf 84}, 024432 (2011).\\
35.  J. J. Hauser, Amorphous concentrated spin-glasses: Mn$X$ ($X=$ Ge, C, Si-Te), Phys. Rev. B {\bf 22}, 2554 (1980); S. H. Song, M. H. Jung and S. H. Lim, Spin Glass Behavior of Amorphous Ge-Mn Alloy Thin Films, J. Phys. Condens. Matter {\bf 19}, 036211 (2007); L. Zeng, J. X. Cao, E. Helgren, J. Karel, E. Arenholz, L. Ouyang, D. J. Smith, R. Q. Wu and F. Hellman, Distinct Local Electronic Structure and Magnetism for Mn in Amorphous Si and Ge, Phys. Rev. B {\bf 82}, 165202 (2010).\\
36.  H. Maletta and W. Felsch, Insulating Spin-Glass System Eu${_x}$Sr$_{1-x}$S, Phys. Rev. B {\bf 20}, 1245 (1979).\\
37.  Suggested by V. Martin-Mayor, private communication.\\
38.  W. H. Press, S. A. Teukolsky, W. T. Vetterling, and B. P. Flannery, {\it Numerical Recipes in C: The Art of Scientific Computing}, Second Edition, Cambridge  University Press, Cambidge (1992), Chapters 6 and 15.\\
39.  S. Franz, G. Parisi and M. A. Virasoro, Interfaces and lower critical dimension in a spin glass model, J. Phys. I {\bf 4}, 1657 (1994).\\
40.  S. Boettcher, Stiffness of the Edwards-Anderson Model in all Dimensions, Phys. Rev. Lett. {\bf 95}, 197205 (2005).\\
41.  L. W. Lee and A. P. Young, Large-scale Monte Carlo simulations of the isotropic three-dimensional Heisenberg spin glass, Phys. Rev. B {\bf 76}, 024405 (2007).\\
42.  L. A. Fernandez, E. Marinari, V. Martin-Mayor, G. Parisi and J. J. Ruiz-Lorenzo, Universal Critical Behavior of the Two-Dimensional Ising Spin Glass, Phys. Rev. B {\bf 94}, 024402 (2016).

\end{document}